# Using Multi Expression Programming in Software Effort Estimation

Najla Akram, AL-Saati, Taghreed Riyadh Alreffaee

*Software Engineering Dept., University of Mosul, Iraq*

*Abstract* - Estimating the effort of software systems is an essential topic in software engineering, carrying out an estimation process reliably and accurately for a software forms a vital part of the software development phases. Many researchers have utilized different methods and techniques hopping to find solutions to this issue, such techniques include COCOMO, SEER-SEM,SLIM and others. Recently, Artificial Intelligent techniques are being utilized to solve such problems; different studies have been issued focusing on techniques such as Neural Networks NN, Genetic Algorithms GA, and Genetic Programming GP. This work uses one of the linear variations of GP, namely: Multi Expression Programming (MEP) aiming to find the equation that best estimates the effort of software. Benchmark datasets (based on previous projects) are used learning and testing. Results are compared with those obtained by GP using different fitness functions. Results show that MEP is far better in discovering effective functions for the estimation of about 6 datasets each comprising several projects.

## I. INTRODUCTION

With the recent rapid advance in area of software engineering, software have become widely and reliably used in business corporations and manufacturing industrial sectors. This is due to its accuracy and efficiency in achieving the required tasks. The use of such packages forms both a source of progression and an economical benefit to various organizations by saving time and cost.

Accurate estimation of size, cost, effort, and time tables for any software denotes the major challenge facing software developers nowadays. Its main influence on the software development management resides in the direct impact of both underestimation and overestimation in causing damages to software foundation's [1-2]. Estimating software effort or cost precisely can provide successful planning for project managers who significantly reduce the risk of making uncertain decisions about project activities and accountabilities.

The importance of achieving accurate effort estimation for software has always attracted researcher to seek methods that can accomplish such accuracy, algorithmic methods such as COCOMO [3], Putnam model [4], and function points based models [5], are in general incapable of dealing with exceptional conditions, particular experience and factors are not easily quantified, as well as the inaccuracy of cost driver rating that can lead to imprecise estimation. Non-algorithmic methods, on the other hand, are more flexible to use and provide the incorporation of human intelligence and their intuitive experience to help achieve estimations that are more reliable. These methods relate entirely too computational intelligence methods such as Genetic Algorithm GA, Neural Networks NN, Fuzzy Logic, and Swarm Intelligence.

In this work a study is introduced to show the possibility of applying one of the linear Genetic Programming GP methods, aiming at providing a function capable of yielding as accurate as possible estimation of software effort. This method is called Multi Expression Programming [6], a linear variation of GP used to reduce the complexity experienced with traditional GP by eliminating dealing with trees and linked lists, handling and encoding chromosomes linearly. Results are compared with GP using various benchmark datasets.

## II. RELATED STUDIES

Various research articles have been introduced in the field of software reliability, each employing a different methodology leading to variations in the gained results, some of these are:

In 2001, Dolado [7] employed GP to find a function that calculates the cost; results were compared with other pervious outcomes. Lefley and Shepperd [8], in 2003 similarly explored the use of GP in improving the process of software effort estimation based on general sets of data. In 2004, Ohsugi, et al. [9] proposed a method for effort estimation based on Collaborative Filtering and retrieval of lost data as a strategy of estimation using Defective Data.

GA was used in 2006 by Huang and Chiu [10] to measure software effort via unequal weights, linear and non-linear weights. In 2008, the idea of Bayesian Network Models was used by Mendes and Mosley [11] in a comparative study for web cost estimation. Sheta and Al-Afeef [12] in 2010 used GP to evolve a mathematical model for effort estimation using two variables (Methodology and LOC) in order to evolve a relationship between them.

In 2012, Ziauddin, Tipu, and Zia [13] found a model for estimating the effort of Agile Software Projects using traditional methods and test data of 21 projects. Arnuphaptrairong [14] in 2013 proposed the use of Function Point FP with Data flow Diagram to solve the problem of gaining estimation information in early stages of software development, as most of the estimation models were dependent on information gained in the last stages of development.

Lately in 2015, Ruchi Puri and Iqbaldeep Kaur [15] presented a Novel Meta-Heuristic Algorithmic Approach to estimate software cost. They presented BAT algorithm and Human Opinion Dynamics algorithms for cost estimation using effort parameter.

Recentlyin 2016, Shivani Sharma, Aman Kaushik, and Abhishek Tomar [16] used a Hybrid Algorithm to solve the software cost estimation problem; their objective was to compute the budget of the project based on a Top down method that included computing the function points of each module.

### III. EFFORT ESTIMATION

In the field of software engineering, effort is definable as the total time spent by members of the development team to accomplish the required task. It is usually stated in terms of man- day, man-month, or man-year. There are many reasons tomotivate the estimation of effort such as [17]:

- **Project Approval**. Deciding the launch of a project on the part of an organization, preceded by estimation of effort needed for positive project completion.
- **Project Management**. Managers plan and manage projects, which in turn require estimation of effort as per respective phases so as to finalize a project.
- **Development team members understanding**. For the development team to perform professionally, its members have to understand their specific roles along with the total activities of the team.
- **Project task definition**. This can be done using effort estimation.
- **Accuracy of effort estimation**. This has formed an important subject to researchers for the past 25 years. Various works categorize effort estimation methods variably. Classifications taken into consideration are:
  - Empirical Parametric (Algorithmic) estimationmodels;
  - Empirical non-parametric estimation models;
  - Expert estimation;
  - Analogue estimation models;
  - Downward estimation;
  - Upward estimation.

Several independent surveys have been carried out for the importance of effort estimation in the area of software development; these investigations showed that 70-85% of respondents agreed on the importance of effort estimation. [2][18]Effort estimation methods can be classified into the following [19]:

- **Historical Analogy:** when similar previous historical data, this data (registered, recorded, associated with previously completed projects) can be used to calculate the effort for future projects.
- **Experts'Decision:** estimating effort this way usually depend on a human expert, the expertise of a human depends largely on how similar are his previously faced projects with the currently required to be estimated project. This method is fairly accuratewhen the estimator has enough experience in both software and estimation.

- **The use of models:** this can include estimations created using mathematical or parametric cost models. Such equations have been derived essentially by means of statistical methods; they usually involve human effort, cost and schedule.
- **Rules-of-thumb:** such rules may differ from simple mathematical equations to specifying a percentage of the activities or phase's effort depending on pervious historical data.

In the field of system engineering, pervious historical data is considered as a source for estimating future effort or cost. Unfortunately in the field of software production, it is frequently very hard, if not impossible, to find reliable datasets.

At the phase of design and construction for a project, the process of estimating the effort is considered to be very hard and complicated for the following reasons [20]:
- A project of this size or type has never been built before.
- Some new techniques are employed in it, which has never been used earlier.
- The Productivity of personnel is largely inconsistent.

The COCOMO model [3] is one of the first methods used in calculating the effort automatically, where the estimated effort is a function of expected size as stated in Eq.(1).

$$E = aS^b \quad \quad \quad \quad \quad \quad \quad \quad \quad (1)$$

Where
E: is the required effort.
S: is the expected size.
a, b: are constants.

The advantages of using COCOMO are [21]:
- It is easy to adapt and is very understandable.
- It provides more objective and repeatable estimations.
- It creates the possibility of calibrating the model to reflect any type of software development environment and thus, providing more accurate estimates.
- Works on historical data and hence is more predictable and accurate.

While some of the disadvantages found in the COCOMO model are [21]:

- This model ignores requirements and all documentation.
- It ignores customer skills, cooperation, knowledge and other parameters.
- It oversimplifies the impact of safety/security aspects.
- It ignores hardware issues
- It ignores personnel turnover levels
- It is dependent on the amount of time spent in each phase.

Most models of effort estimation relay on Empirical Derivation using pervious project's data, where the software size is the input to the calculation. This size is measured using LOC or FP. [22]

### IV. GENETIC PROGRAMMING

The concept of Genetic Programming is derived from the well-known idea of Genetic Algorithms in a trial to answer one of the basic questions in computer science: [23]

*"How can computers learn to solve problems without being explicitly programmed? In other words, how can computers be made to do what is needed to be done, without being told exactly how to do it?"*

Genetic programming is a way of generating computer programs automatically contributing vary effectively in solving carefully specified problems, forming one of Evolutionary Computational techniques. This approach was successfully used to solve a huge number of difficult problems such as modeling industrial operations, water flow prediction and others. [12]

Genetic Programming is one of the evolutionary algorithms based on the evolutionary theory and the nature's survival of the best idea. These algorithms depend on forming a population of individuals each represented as trees expressing an equation or a program where there is no constraint on the resulting data structure. [24]

There are four steps needed to establish GP, they are necessary to solve the problem at hand [25]:
- Define Terminal and Function sets as stated by the problem.
- Set the appropriate fitness function according to the problem specification.

- Set Control Parameters including (iteration number, tree size and depth, population size, and crossover and mutation rates, etc.)
- Set stopping criteria.

After that, chromosomes of the first population are randomly generated; each consisting of a random combination of variables and functions appropriate for the problem at hand, these variables and functions are usually predefined by the user. The chromosomes unlike GA have varying lengths in the population as each encoded equation or program consists of a different number of variables and functions.

The fitness of each chromosome in the population is evaluated using a fitness function to estimate the chromosome's behavior and efficiency. This function is measured in various ways such as error ratio between the actual desired input and the achieved output. It may also be measured through the required (time, cost, or fuel) needed to reach the desired goal. The function can similarly be calculated using the resulting precision of applications such as pattern recognition or object classification according to given problem. [25]

Subsequently, a selection process is conducted in order to specify which of the chromosomes are to be chosen for reproduction and genetic operators, with the intention of forming the offspring for the new generation. This selection is basedon the fitness of individuals, the more fit an individual is the more chance it has to be selected.

## V. MULTI EXPRESSION PROGRAMMING

GP is considered to be very complicated to program and cope with due to the complexity associated with tree structures. Thus many linear variations have been proposed in the literature all aiming to simplify the encoding of chromosomes in a linear effective representation. One of these methods is the Multi Expression Programming (MEP). [6]

Multi Expression Programming is a technique that automatically generates computer programs, mathematical expressions, and equations. It is very much the same as GP, the differences residing between them are [26]:

- In GP, each chromosome encodes a single expression. On the other hand, a chromosome in MEP encodes several expressions. Every one of the encoded expressions can be selected to represent the chromosome.
- The encoding of chromosomes is linear in MEP unlike GP, where chromosomes are encoded nonlinearly (as trees).

MEP algorithm begins by initializing the first population randomly, after that a repetition of steps is conducted until a stopping criterion is meat. In each iteration, two parents are selected for recombination producing two offspring; these may be subjected to mutation. The best resulting offspring will replace the worst individual in the population if its fitness was better. In the end, the resulting best individual will carry the best expression developed through the previously specified number of generations.[26] the main steps of MEP are given in Fig. 1.

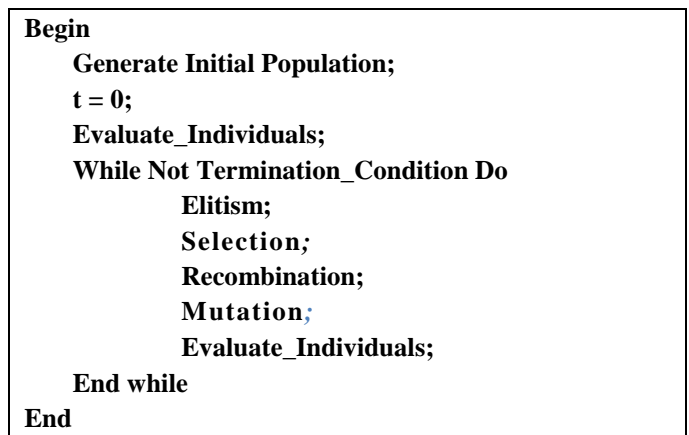

```
Begin
    Generate Initial Population;
    t = 0;
    Evaluate_Individuals;
    While Not Termination_Condition Do
            Elitism;
            Selection;
            Recombination;
            Mutation;
            Evaluate_Individuals;
    End while
End
```

Fig. 1 Basic steps of MEP [6]

### A. Chromosomal Encoding:

Every chromosome has a fixed number of genes, every gene in encoded with either a Terminal symbol or a Function symbol. Genes encoded with a function must contain a pointer to the arguments of that function, and the first gene of the chromosome must always be Terminal.

One of the most important properties of this method is its ability to store multiple solutions to the problem in the same chromosome, with the best solution being chosen according to the fitness. [26]

Assuming a chromosome (c) consisting of multiple genes, a set of functions F={+,*}, and a set of terminals T={a,b,c,d}. Then the encoding of the chromosome according to MEP is:

1: a
2: b
3: + 1, 2
4: c
5: d
6: + 4, 5
7: * 3, 5
8: + 2, 6

The maximum number of a chromosome's symbols is calculated as in Eq.(2).

*No. of Symbols= (n+1)\*(No. of Genes-1)+1 ......... (2)*
Where
n: is the maximum number of arguments taken by a function in set F.

Genes {1,2,4,5} were encoded with simple Expressions, while the rest were encoded with Complex Expression containing functions. Expression 3 employed the function {+} with two operands present as pointer to locations of expressions numbered {1} and {2} in the chromosome. So decoding the third expression will result (E3 = (a + b)) and so on for the rest of the expressions.

**E6 = c + d**
**E7 = (a + b) \* d**
**E8 = b \* (c + d)**

The name of this algorithm, MEP, comes from the fact that it allows encoding multiple expressions and their number is equal to the length of the chromosome (no. of genes). It is noticed that in this chromosome the length is 8 and there exists an equal number of expressions, the final form of the chromosome is:
**E1 = a,**
**E2 = b,**
**E3 = a + b,**
**E4 = c,**
**E5 = d,**
**E6 = c + d,**
**E7 = (a + b) \* d,**
**E8 = b \* (c + d)**

Fig.2 shows the representation of the explained above chromosome (c) as a tree, the numbers underlying each branch are the expression's numbers.

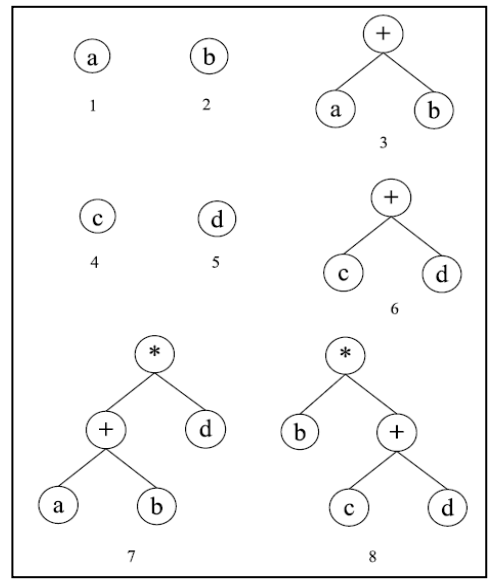

Fig. 2 Representation of Chromosome (c) as a tree [26]

*B. Fitness Function*

In any population, individuals are chosen according to how well they perform on getting closer to reach the required solution; this performance is called the fitness of an individual needed to direct evolution in favor of the best. This fitness is measured in various ways according to the given problem.

One way to evaluate fitness isby measuring the difference between the result of expression $E_i$ called ($O_{k,i}$) and the actual output ($W_k$) both for fitness case (k), this is done as in Eq.(3), here the fitness has to be minimized. After that, the fitness for the individual will be the lowest fitness of the expressions encoded in the chromosome, as in Eq.(4).[26]

$$f(Ei) = \sum_{k=1}^{n} | O_{k,i} - W_k | \ ..................................... (3)$$
$$f(C) = \min f(E_i) \ ..................................... (4)$$

The main feature of MEP resides in overcoming the various problems of GP such as the difficulty of dealing with tree structures and the corresponding effort of applying the genetic operators. In addition to that, there is the problem of predefining tree size and depth, which resembles a very critical problem in the success of GP along with keeping that size in range after successive crossovers between tree branches and mutations and insuring that the resulting program is always correct.

## C. Genetic Operations:

MEP uses the same traditional genetic operators introduced by GA, such as[6]:

- **Recombination (crossover)**: after two chromosomes are being selected using Roulette Wheel or Tournament selection, crossover is implemented according to its probability using:
  - **One-point Recombination**
  - **Two-point Recombination**
  - **Uniform Recombination**

- **Mutation:** every symbol in the chromosome is subjected to the probability of mutation. When a symbol mutates from a terminal to a function, its operands will be automatically generated and when a function mutates to a terminal, its operands are ignored.

## D. Selection:

Selection is the process where individuals are chosen from the population to undergo genetic operations according to their fitness. In this work tournament selection is used to choose two chromosomes randomly from the population to go through tournament. The fitter individual of the competitors will win the tournament and be subjected to genetic operators. This method of selection has many benefits such as being easy to code and program, as well as the possibility of implementation in parallel architectures [27].

The selection process usually guarantees giving a better chance to the more fit individuals in the population to move on to the next generation. [6] Most of the studies have proved through tests and experiments that the best tournament size is (2), this size was used in the experimental part of this work.

## VI. EXPERIMENTAL TESTING AND RESULTS

### A. Datasets

In this work, an investigation has been carried out to show the possibility of finding an estimation function for software effort though out the use of MEP using the Datasets shown in TABLE I.

The chosen Datasets were selected to provide variety and diversity, and due to their availability and recurrent use, they have become benchmark datasets in this field of study used mainly in comparisons among different methods and techniques introduced to estimate software effort.

TABLE I
Data sets used in this work

| No. | Dataset Name | Author's Name | Total no. of Projects |
|-----|--------------|---------------|------------------------|
| 1. | Albrecht & Gaffney[1] | A.J. Albrecht, J.R. Gaffney | 5 incomplete (3,6,7,22,24) 24 points |
| 2. | Bailey & Basili [28] | J.W. Bailey, V.R. Basili | 18 points |
| 3. | Heiat & Heiat [29] | A. Heiat, N. Heiat | 35 points |
| 4. | Kemerer [30] | C.F. Kemerer | 15 points |
| 5. | Miyazaki et. al. [31] | Y. Miyazaki, M.Terakado, K. Ozaki, H. Nozaki | 48 points |
| 6. | Desharnais[32] | J.M. Desharnais | 4 incomplete (38,44,66,75) 77 points |

Next are the experiments carried out along with the results, in addition to the analysis and discussion.

### B. Implementing MEP:

The first test in this work involves the implementation of MEP on the Datasets mentioned in TABLE I. Results are afterwards compared to those obtained by Dolado [7] using GP. A crossover rate of (0.7) and a mutation rate of (0.05) are used though out this experiment. The preparation of the algorithm includes defining the parameter settings as follows:

**Population size:** 40
**Generations:** 200
**Function Set:** {-, +, *, /, POWER, EXP, LOG, SQRT}
**Terminal Set:** The project's variables depending on the Dataset.

TABLE II shows the comparison between MEP's results and those found by GP. Results signify the efficiency of MEP, as all the gained values were noticeably better for all datasets. The best values are shown for the fitness and generation numbers needed to achieve that fitness.

TABLE II
A Comparison between MEP and GP

| No. | Dataset | GP | MEP | |
|---|---|---|---|---|
| | | Fitness | Fitness | Gen. |
| 1. | Albrecht & Gaffney | 0.548 | **0. 33910** | 56 |
| 2. | Bailey & Basili | 0.269 | **0.14200** | 45 |
| 3. | Desharnais | 0.623 | **0.38951** | 67 |
| 4. | Heiat & Heiat | 0.087 | **0.08570** | 100 |
| 5. | Kemerer | 0.584 | **0.36854** | 200 |
| 6. | Miyazaki | 0.506 | **0.32420** | 200 |

*C. Additional Investigation:*

To further examine the efficiency of MEP, a deeper investigation is conducted to show the impact of population size and generations needed to reach the required solution using the same function and terminal sets. This investigation is done using two tests:

- **TEST1**: included small narrowed samples of population size: (10, 20, 30, and 40) as well as for generations: (25, 50, 75, 150, and 250).

- **TEST2**: involved considering larger and more wide apart samples for population sizes and generation numbers: (50, 100, 150,200,250,300,350,400,450,and 500)

Fig.(3−8) illustrate the results of applying the first phase samples on the datasets in TABLE I. Fitness values are shown against Population sizes across colored bars reflecting the generation number (shown in the legend on the right of each graph).

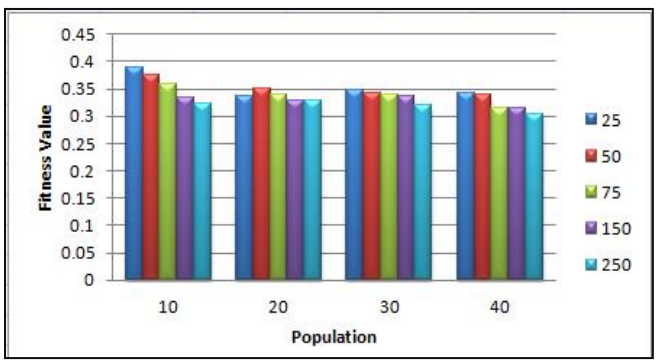
Fig .3Fitness values for TEST1 (Albrecht & Gaffany)

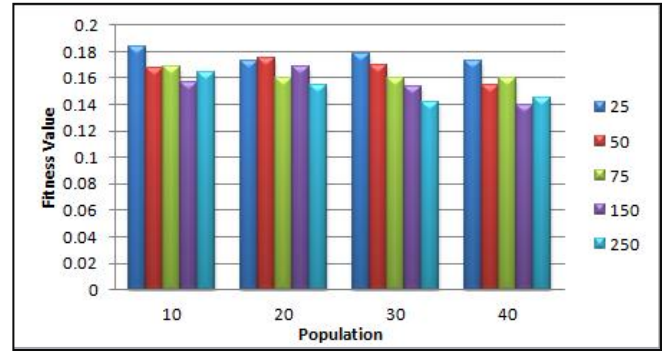
Fig.4 Fitness values for TEST1 (Bailey & Basili)

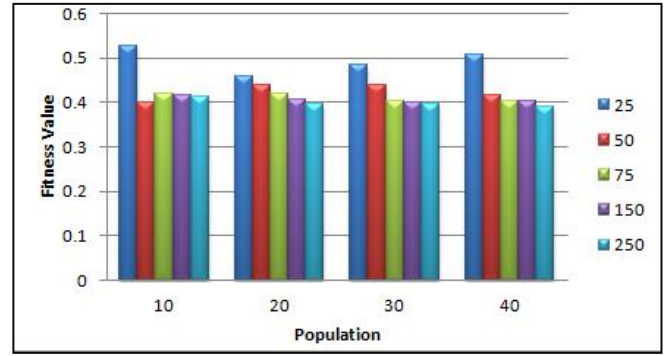
Fig.5 Fitness values for TEST1 (Desharnais)

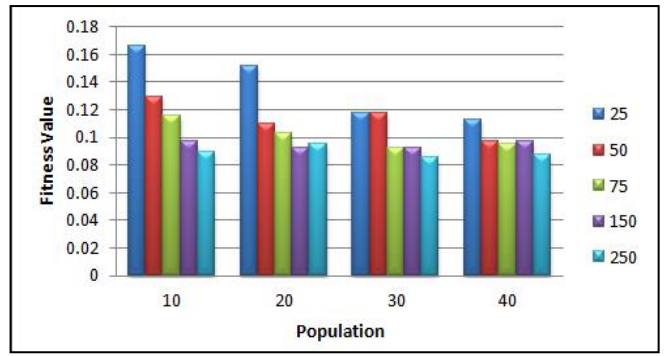
Fig.6 Fitness values for TEST1 (Heiat & Heiat)

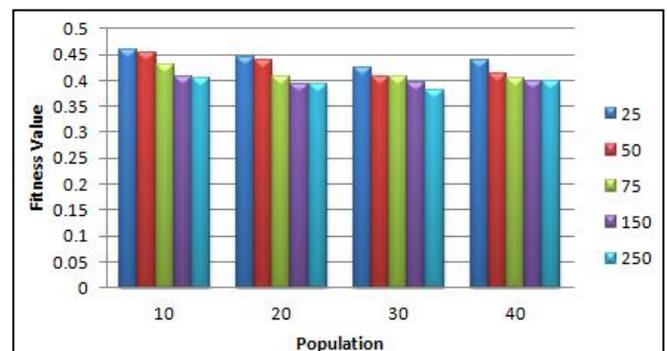
Fig.7 Fitness values for TEST1 (Kemerer)

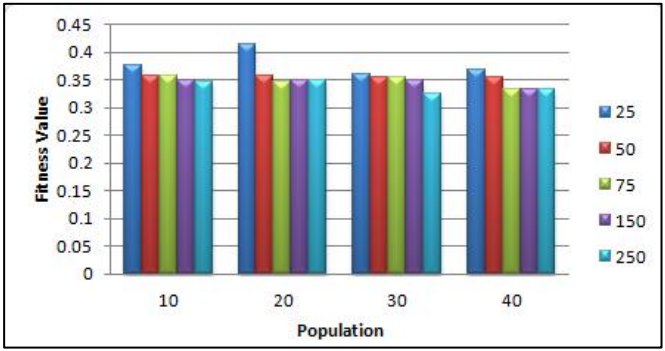

Fig. 8 Fitness values for TEST1 (Miyazaki)

As can be clearly seen from the former figures, MEP has the ability to accomplish a very distinguished success with very small population sizes and of generations' number. They also confirm the fact that increasing generations raises theprobability of gaining better results. It is also obvious that theexpansion of population sizes does not have much impact on achieving better results (less fitness values).

On the other hand, Fig. (9 - 14) demonstrate the fitness values for larger population sizes and greater number of generations as given in TEST2 in an attempt to traverse a wider area of the search space for the problem, and to investigate the strategy of the algorithm in searching for better results and more suitable ones for the employed datasets.

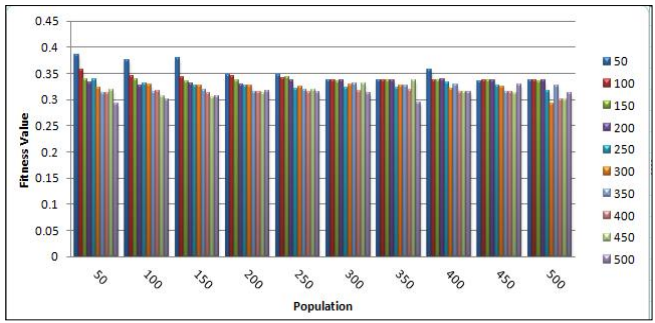

Fig.9 Fitness values for TEST2 (Albrecht & Gaffany)

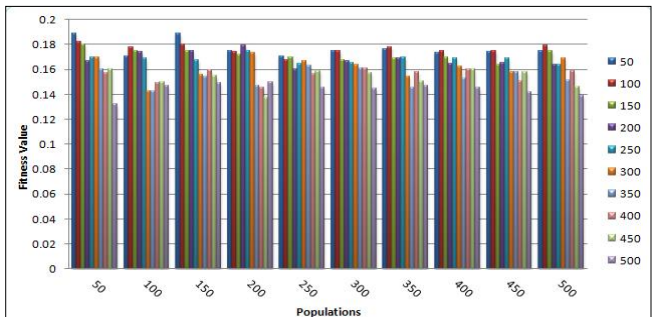

Fig.10 Fitness values for TEST2 (Bailey & Basili)

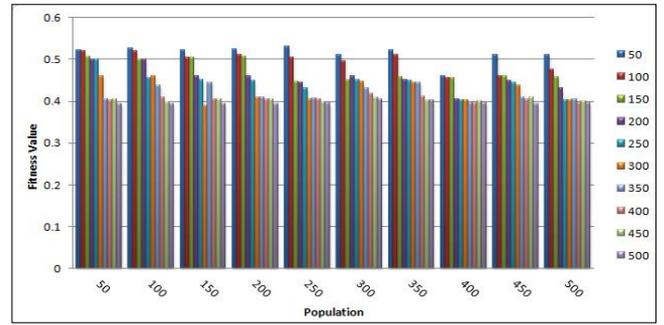

Fig. 11 Fitness values for TEST2 (Desharnais)

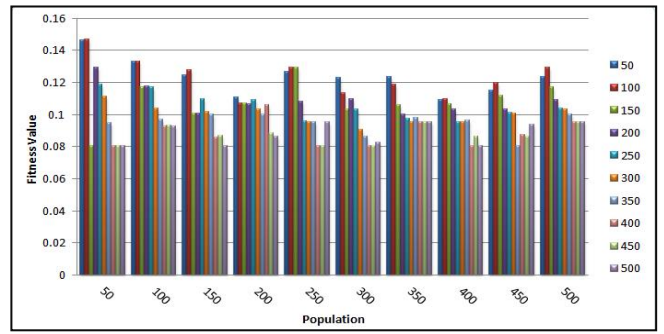

Fig. 12 Fitness values for TEST2 (Heiat & Heiat)

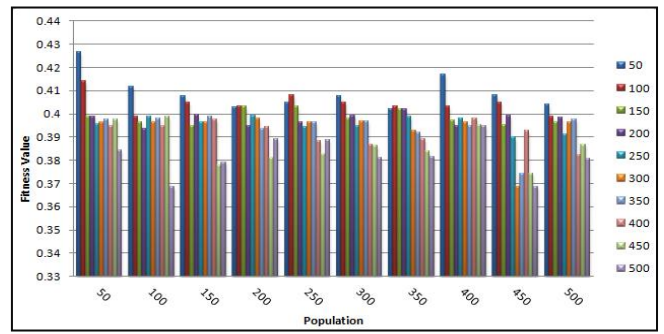

Fig. 13 Fitness values for TEST2 (Kemerer)

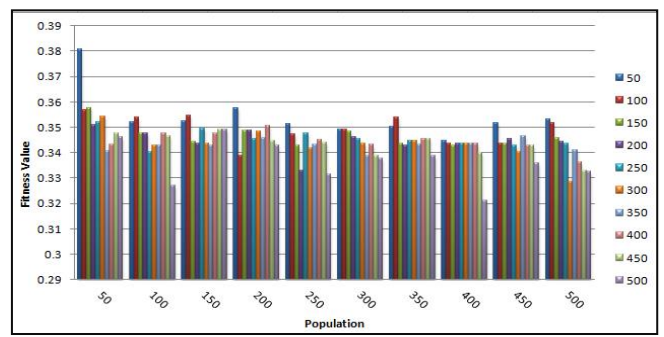

Fig. 14 Fitness values for TEST2 (Miyazaki)

This investigation indicates that increasing generations allow for better solutions in general for all datasets. But then again, larger population sizes did not have an effect on improving fitness; this indicates that small population sizes taken in TEST1 were sufficient enough to achieve the same results.

As a whole, investigating a wider search space for problems did not improve the results significantly, but it has helped in establishing the appropriate sizes required by the evolutionary algorithm to reach good enough solutions and far better than those gained using GP as shown in TABLE II.

The efficiency of MEP verified in this investigation is related to the underlying structure of the chromosome, as it encodes multipleexpressions instead of only one. This structure increases the effectiveness of the resulting solution.Fig. 15 depicts a sample solution exemplifying the obtained equation for effort estimation.

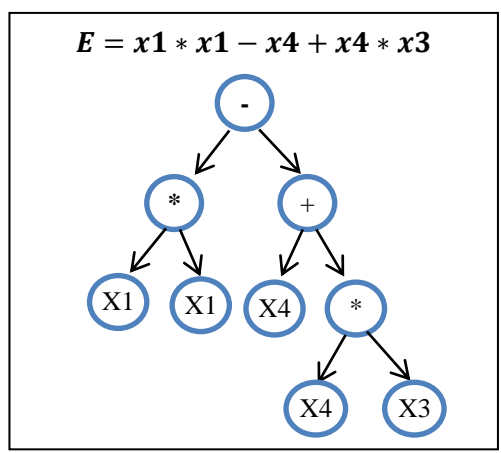

Figure (15) A sample solution represented by a tree

## VII. CONCLUSIONS

The main purpose of this work is to efficiently adopt the intelligence found in Artificial Intelligent Techniques such as Genetic Programming in finding effort calculating equations to estimate software effort. This was done using MEP algorithm, one of the GP linear variants in order to overcome the difficulties of coding GP and application of genetic operators on trees, not forgetting the obstruction of predefining tree size and depth and keeping solutions correctly functioning after crossover and mutation.

MEP has been applied successfully in this work to solve the software effort estimation problem, the algorithm was able to come up with very satisfying solutions encoded in correctly formulated chromosomes. These solutions (equations) are capable of giving an estimation of effort for projects before its establishment, and thus help in getting such a project completed efficiently and satisfactorily. Results were compared with those obtained by GP and found to be far precise and accurate.

In addition, an investigation was performed to show the impact of different population sizes and varying generation numbers on fitness values, this has proven that large populations did not have an effect on providing better results; this is due to the efficiency of the algorithm employed. Higher generation numbers, on the other hand, had an impact, although not so significant, on refining the fitness values obtained.

## VIII. REFERENCES

[1] A.J.Albrecht1,J.R. Gaffney, (1983)," *Software Function, Source Lines of Code, and Development Effort Prediction: a Software Science Validation*", IEEE Trans. on SWE. 9(6) PP:639–648.

[2] R.Bhatnagar, M.K.Ghose, (2012) "*Early Stage Software Development Effort Estimations-Mamdani FIS VS Neural Network Models*". CS&IT.pp:377–384.

[3] B.W. Boehm, (1981) *Software engineering economics*, Englewood Cliffs, NJ: Prentice-Hall.

[4] L. H. Putnam, (1978)."*A general empirical solution to the macro software sizing and estimating problem*". In IEEE Trans.on SWE, 4(4), 345-361.

[5] S.A.Abbas, et. al. (2012) "*Cost Estimation: A Survey of Well-known Historic Cost Estimation Techniques*", J. of Emerging Trends in Computing and Information Sciences, 3(2), pp. 612-636.

[6] M.Oltean, D.Dumitrescu,(2002)."*MultiExpression Programming*".TechnicalReport,UBB-01-2002.

[7] J.J.Dolado, (2001) . "*On the problem of the software cost function*", Information and Software Technology, p:2001 Elsevier Science B.V.

[8] M.Lefley, M.Shepperd, (2003), "*Using genetic programming to improve software effort estimation based on general data sets*". In Procs. of Genetic & Evol. Comp. Conference, 2003, 2477–2487.

[9] N.Ohsugi, M.Tsunoda, A.Monden, and K.Matsumoto, (2004),"*Effort Estimation Based on Collaborative Filtering*", In the 5th International Conference on Product Focused Software Process Improvement (PROFES2004), pp. 274-286.

[10] S.Huang, N.Chiu,(2006) "*Optimization of Analogy Weights by Genetic Algorithm for Software Effort Estimation*".J. of Sys.& SW 48(11), pp:1034-1045.

[11] E.Mendes, N.Mosley, (2008). "*Bayesian Network Models for Web Effort Prediction: A Comparative Study*". IEEE Trans. SWE, 34(6), pp: 723-737.

[12] A.F.Sheta, A.Al-Afeef, (2010). "*A GP Effort

*Estimation Model Utilizing Line of Code and Methodology for NASA Software Projects*", In proc. of 10th International Conf. on Intelligent Systems Design and Apps., ISDA, pp: 290-295.
[13] Sh. Ziauddin, T. Kamal, Z. Shahrukh, (2012) "*An Effort Estimation Model for Agile Software Development,*" Advances in Computer Science and Its Applications (ACSA), Vol.2, No.1, pp. 314-324.
[14] T.Arnuphaptrairong, (2013),"*Early Stage Software Effort Estimation Using Function Point Analysis: Empirical Evidence*", Proc. of the Inter. Multi-Conf. of Engineers and Computer Scientists Vol. II, (IMECS), March 13-15, Hong Kong. pp: 730-735.
[15] R. Puri, I. Kaur, (2015) "*Novel Meta-Heuristic Algorithmic Approach for Software Cost Estimation*". In I. J. of Innovations in Engineering and Technology (IJIET), Vol(5), Issue-2.
[16] Sh. Sharma, A. Kaushik, and A. Tomar, (2016) "*Software Cost Estimation using Hybrid Algorithm*". In I. J. of Engineering Trends and Technology (IJETT). Vol.(37), No.2.
[17] J.Živadinović, Z.Medić, D.Maksimovi,A.Damnjanović, S.Vujčić, (2011) "*Methods Of Effort Estimation In Software Engineering*", In Inter. Symp. Eng. Manag.& Competitiveness (EMC2011), June 24-25, Zrenjanin, Serbia.
[18] I.Z., Quba, (2012). "*Software Projects Estimation using Neural Networks*". M.Sc. Thesis. College of Computers Sciences &Math. University of Mosul.
[19] A.Tsakonas, G.Dounias, (2009). "*Deriving Models for Software Project Effort Estimation by Means of Genetic Programming*". In KDIR-2009 Workshop (INSTICC),6-8 October, Madeira, pp: 34-42.
[20] J.Asundi, (2005),"*The Need for Effort Estimation Models for Open Source Software Projects*", Software Engineering (5-WOSSE) May 17,St Louis, MO, USA. ACM 1-59593-127-9.pp:1-3.
[21] P. Rijwani, S. Jain and Dh. Santani, (2014). "*Software Effort Estimation: A Comparison Based Perspective.*" In I. J. of App. or Innovation in Eng.& Manag. (IJAIEM). Vol.(3), Issue-12,. pp:18-29.
[22] B.Jeng, D. Yeh, D. Wang, S.Chu, and C.Chen, (2011),"*A Specific Effort Estimation Method Using Function Point*", Journal Of Information Science And Engineering 27, pp: 1363-1376.
[23] J.R. Koza, (1994), "*Genetic Programming II: Automatic Discovery of Reusable Programs*". ©1994 Massachusetts Institute of Technology.
[24] J.R.Koza, (1992),"*Genetic Programming: On the Programming of Computers by Means of Natural Selection*",©1992Massa. Institute of Technology.
[25] J.R.Koza, M. A.Keane, M. J.Streeter, W.Mydlowec, J.Yu, G. Lanza, (2003) "*Genetic Programming IV Routine Human-Competitive Machine Intelligence*", ISBN 1-4020-7446-8. © 2003 Springer Science+Business Media, Inc.
[26] M.Oltean,(2006),"*Multi Expression Programming*". Tech.l Report, Babes-Bolyai Univ, Romania.28p.
[27] B.L.Miller; D.E.Goldberg, (1995). "*Genetic Algorithms, Tournament Selection, and the Effects of Noise*". Complex Systems. 9: 193–212.
[28] J.W.Bailey, V.R.Basili, (1981), "*A Meta-model for Software Development Resource Expenditures*". Proc. of the 5th Inter. Conf. on SWE, pp: 107–116.
[29] A.Heiat, N.Heiat, (1997) "*A Model for Estimating Efforts Required for Developing Small-Scale Business Applications*", In Journal of Systems and Software 39 (1) pp:7–14.
[30] C.F.Kemerer, (1987), "*An Empirical Validation of Software Cost Estimation Models*", Comm. of the Association for Comput. Machin.30(5).pp:416–429.
[31] Y.Miyazaki, M.Terakado, K.Ozaki, H.Nozaki, (1994), "*Robust regression for developing softwareestimation models*", J. of Sys.& SW 27(1),pp:3–16.
[32] J.M.Desharnais, (1988), "*Analyse statistique de la productivite´ des projects de de´velopment en informatique a` partir de la technique des points de function*", M.Sc. Thesis, Univ. du Que´bec a` Montreal, De´cembre,.